# Hubble Space Telescope Spectroscopy of the Balmer lines in Sirius B[†]

M.A. Barstow[1], Howard E. Bond[2], J.B. Holberg[3], M.R. Burleigh[1], I. Hubeny[4] and D. Koester[5]

[1]*Department of Physics and Astronomy, University of Leicester, University Road, Leicester LE1 7RH UK*
[2]*Space Telescope Science Institute, 3700 San Martin Dr., Baltimore, MD 21218 USA*
[3]*Lunar and Planetary Laboratory, University of Arizona, Tucson, AZ 85721 USA*
[4]*Steward Observatory, University of Arizona, Tucson, AZ 85721 USA*
[5]*Institut für Theoretische Physik und Astrophysik, Universität Kiel, 24098 Kiel, FRG*

25 May 2005

**ABSTRACT**

Sirius B is the nearest and brightest of all white dwarfs, but it is very difficult to observe at visible wavelengths due to the overwhelming scattered light contribution from Sirius A. However, from space we can take advantage of the superb spatial resolution of the Hubble Space Telescope to resolve the A and B components. Since the closest approach in 1993, the separation between the two stars has become increasingly favourable and we have recently been able to obtain a spectrum of the complete Balmer line series for Sirius B using *HST*'s Space Telescope Imaging Spectrograph (STIS). The quality of the STIS spectra greatly exceed that of previous ground-based spectra, and can be used to provide an important determination of the stellar temperature ($T_{eff}$ = 25193K) and gravity (log $g$ = 8.556). In addition we have obtained a new, more accurate, gravitational red-shift of 80.42 ± 4.83 km s$^{-1}$ for Sirius B. Combining these results with the photometric data and the *Hipparcos* parallax we obtain new determinations of the stellar mass for comparison with the theoretical mass-radius relation. However, there are some disparities between the results obtained independently from log $g$ and the gravitational redshift which may arise from flux losses in the narrow 50x0.2″ slit. Combining our measurements of $T_{eff}$ and log $g$ with the Wood (1995) evolutionary mass-radius relation we get a best estimate for the white dwarf mass of 0.978 M$_\odot$. Within the overall uncertainties, this is in agreement with a mass of 1.02 M$_\odot$ obtained by matching our new gravitational red-shift to the theoretical *M/R* relation.

## 1 INTRODUCTION

As the nearest and visually brightest example, Sirius B is one of the most important of all the white dwarf stars. Detected by Bessel (1844) through membership of a binary system, with its companion Sirius A, it provides an opportunity for an astrometric mass determination. This can be compared with other independent methods of determining stellar mass from spectroscopic temperature and gravity measurements and an observation of the gravitation redshift. Unfortunately, the proximity of Sirius B to the primary star makes most accurate spectroscopic and photometric observations extremely difficult. For example, at visible wavelengths Sirius A is approximately 10 magnitudes brighter than Sirius B. Only at the shortest far-UV wavelengths or in the EUV/soft X-ray band does Sirius B become brighter than Sirius A. Of course, observations in these wavelength ranges only became possible in the space-age.

Therefore, for most of the time since its discovery astronomers have needed to make the most challenging of observations to learn about Sirius B. During this time few useful spectra of the star were obtained. The first, by W.S. Adams (1915) revealed the enigma of white dwarf stars, showing Sirius B and Sirius A to be "identical in all respects so far as can be judged from a close comparison of the spectra", the much lower luminosity of Sirius B placing it in the lower left corner of the H-R diagram along with 40 Eri B. From subsequent observations Adams (1925) reported a first gravitational redshift. However, along with that reported by Moore (1928), the value was a factor 4 too low due to contamination of the spectra by Sirius A (see discussions by Greenstein, Oke and Shipman 1971, 1985; Wesemael 1985). For example, the results depended on measurements of metallic lines such as MgII 4481Å, which are now known not to occur in most white dwarfs. Indeed, a reliable redshift (89±16km/s) was only eventually published by Greenstein et al. (1971) based on a photographic plate obtained in ~1963.

Greenstein et al. (1971) also published measurements of the effective temperature and surface gravity of Sirius B, based on their analysis of the Hα and Hγ profiles, of 32000±1000K and log $g$ = 8.65, respectively. However, measurements based on only a few such line profiles can be prone to ambiguities in the determination of these parameters. Also, modern computational fitting techniques allow a complete objective exploration of the available parameter space compared to the visual comparisons available to Greenstein et al. (1971). While, the effective temperatures of DA white dwarfs can be formally measured to an accuracy of ~1% between 20000K and 30000K, with access to the complete Balmer line series from Hβ to H8 (see Finley, Koester & Basri. 1997; Liebert Bergeron & Holberg 2005), estimates of $T_{eff}$ for Sirius B have ranged from the Greenstein et al. value down to as low as 22500K (Koester 1979). The availability of low dispersion *IUE* and *EXOSAT* spectra refined the value of $T_{eff}$ to 26000±2000K (Holberg, Wesemael & Hubeny 1984; Paerels et al. 1988; Kidder, Holberg & Wesemael 1989). Most recently, Holberg et al. (1998) have combined the *IUE* NEWSIPS data and the *Extreme Ultraviolet Explorer* (*EUVE*) spectrum to produce a new, well-defined effective temperature of 24790±100K and surface gravity of log $g$ = 8.57±0.06. Coupled with the *Hipparcos* parallax of π = 0".37921±0".00158 and the Greenstein et al. gravitational redshift, Holberg et al obtained a white dwarf mass of 0.984±0.074$M_\odot$ and a radius $R$ = 0.0084±0.00025$R_\odot$. This spectroscopic result is consistent with the astrometric mass of Gatewood and Gatewood (1978) and the joint spectrometric and astrometric constraints are within 1σ of the most recent thermally evolved M-R relations of Wood (1992, 1995). Some of these earlier measurements are summarized in table 2. Nevertheless, the **spectroscopic** uncertainties remain large at 7.5% in mass and 4% in radius. A precise test of the mass-radius relation for a ~1$M_\odot$ white dwarf, that can distinguish between different evolutionary models, for example, between thick and thin H layer, masses, requires significantly reduced errors on the measured spectroscopic parameters. While important improvements were achieved by Holberg et al. (1998), it is clear that the Balmer line technique provides results of potentially greater accuracy if such a spectrum could be obtained free from contamination by Sirius A.

The paper of Greenstein et al. (1971) does not reproduce the original photographic plate images, plotting just the scanned Hα and Hγ profiles. Hence, it is difficult to judge the level of contamination from Sirius A in their work. Nor in any of the earlier work are the original photographic plates reproduced in the literature. Therefore, a photographic spectrum obtained and published by Kodaira (1967) is of particular importance. This plate covers the Balmer line series from Hγ through to H10 and the spectrum of Sirius B is clearly visible in the middle of scattered light contributions from the diffraction spikes of Sirius A. It also illustrates the particular difficulty of observing the Sirius B from the ground since, even with ~1″ "seeing", the spectrum sits on a scattered light component ~1/4 to 1/3 of its total flux.

Clearly, a visible band observation of Sirius B would be much better carried out in space, reducing considerably the problems discussed above. The Hubble Space Telescope was the first instrument capable of obtaining such a spectrum but for some time following its launch Sirius B has been in an unfavourable position relative to Sirius A, making its closest approach (as projected on the plane of the sky) during 1993. However, as the distance between the two binary companions has increased and we have obtained high quality direct images of the system with *HST*, it has become feasible to obtain a Balmer line spectrum of Sirius B. We report here on the analysis of the spectrum acquired with the Space Telescope Imaging Spectrograph (STIS) during 2004, obtaining measurements of $T_{eff}$ and log $g$, a new gravitational redshift and a revised estimate of the visual magnitude, from which we determine the white dwarf mass and radius.

## 2  THE STIS OBSERVATION OF SIRIUS B

Sirius B was observed with STIS on 2004 February 6, using the G430L and G750M gratings, to obtain coverage of the full Balmer line series (see Table 1). Even though *HST* has the advantage of operating above the atmosphere, acquisition of a spectrum uncontaminated by Sirius A remains a challenge, particularly as the length of the spectrograph slit (52") is considerably greater than the dimensions of the Sirius system. As the orbit of Sirius is well determined, it is quite straightforward to avoid placing Sirius A on the slit at the same time as Sirius B. However, it is inevitable in avoiding the primary that its diffraction spikes must then cross the slit and may potentially contaminate the Sirius B spectrum. To reduce the level of contamination in the Sirius B spectrum to lowest possible level, we chose a spacecraft orientation such that the target was equidistant between the locations of the Sirius A diffraction spikes. Although we devised this approach independently, it is interesting to note that this same technique was adopted by Kodaira (1967), although with less freedom available in the slit orientation. This is illustrated in figure 1, which shows our most recent *HST* Wide Field Planetary Camera 2 (WFPC2) image of the Sirius system, obtained as part of an ongoing programme of imaging with which we are continually improving the astrometric orbit determination and, ultimately, the Sirius B astrometric mass. The image shows the overexposed image of Sirius A, with its four diffraction spikes. The vertical bar is the "bleed" of Sirius A into adjacent pixels along the readout columns of the CCD, due to the overexposure. Sirius B lies just to the right of the bottom left diffraction spike. The grey box represents the dimensions of the 52x0.2" slit, which cuts across Sirius B and two of the diffraction spikes. Note that in the spectroscopic exposure, the main Sirius A image is obscured and the vertical overexposed columns would not be present. With relative freedom to choose the roll angle of *HST* during the spectroscopic exposures, we selected an orientation that placed Sirius B almost exactly half way between two of the diffraction spikes from Sirius A, in the direction perpendicular to the bleeding columns.

3  *M.A. Barstow et al.*                                          *HST spectroscopy of Sirius B*

**Table 1.** Details of the STIS observations of Sirius B

| Grating | λλ range (Å) | Resolution (Å) | File ID | Exposure time (s) |
|---|---|---|---|---|
| G430L | 2900-5700 | 5.5 | O8P901010 | 10.5 |
| G750M | 6265-6835 | 1.1 | O8P901020 | 90.0 |

The G430L and G750M spectra were each obtained as a series of three separate exposures to maximize the signal-to-noise, while preventing saturation of the CCD, and for cosmic ray rejection. The CCD image of the G430L observation is shown in figure 2. Uppermost is the image displayed with a linear intensity scale, which shows the well separated spectra of Sirius B (centre) and the fainter Sirius A diffraction spikes (top and bottom). Absorption dips from the Balmer lines can be clearly seen in all thee spectra. Hβ is to the right of the image and the converging series limit to the left. Below this is the same image using a logarithmic intensity scale, which enhances the lower flux levels showing the background and scattered light components. The presence of the Balmer absorption line in the background shows it to be dominated by the scattered light from Sirius A.

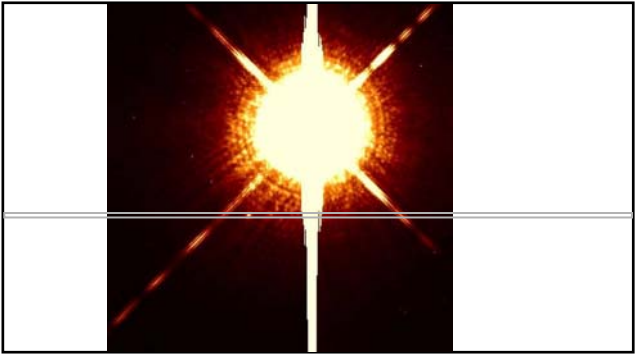

**Figure 1.** *HST* Wide Field Planetary Camera 2 (WFPC2) image of the Sirius system showing the overexposed image of Sirius A, with its four diffraction spikes. The vertical bar is the "bleed" of Sirius A into adjacent pixels along the readout columns of the CCD, due to the overexposure. Sirius B lies just to the right of the bottom left diffraction spike. The grey box represents the dimensions of the 52x0.2" slit, which cuts across Sirius B and two of the diffraction spikes.

To demonstrate the relatively low level of the scattered light, we display a cut through the image in figure 3. Also shown is an equivalent slice through the spectrum of Kodaira (1967), copied from the original paper and normalized to the same intensity and spatial scale as the *HST* data. This illustrates very clearly how the ground-based observation has been compromised by the difficulties with "seeing". In the *HST* image, the diffraction spikes have a much lower intensity than Sirius B and the diffraction limited imaging provides a clear separation of the spectra. Although the scattered light component can be seen in the heavily contrast enhanced lower image of figure 2, it is barely detectable in the intensity histogram. We estimate that the scattered light component is very much less than 1% of the flux of Sirius B in the G430L observation and is ~2% in the G750M grating exposure.

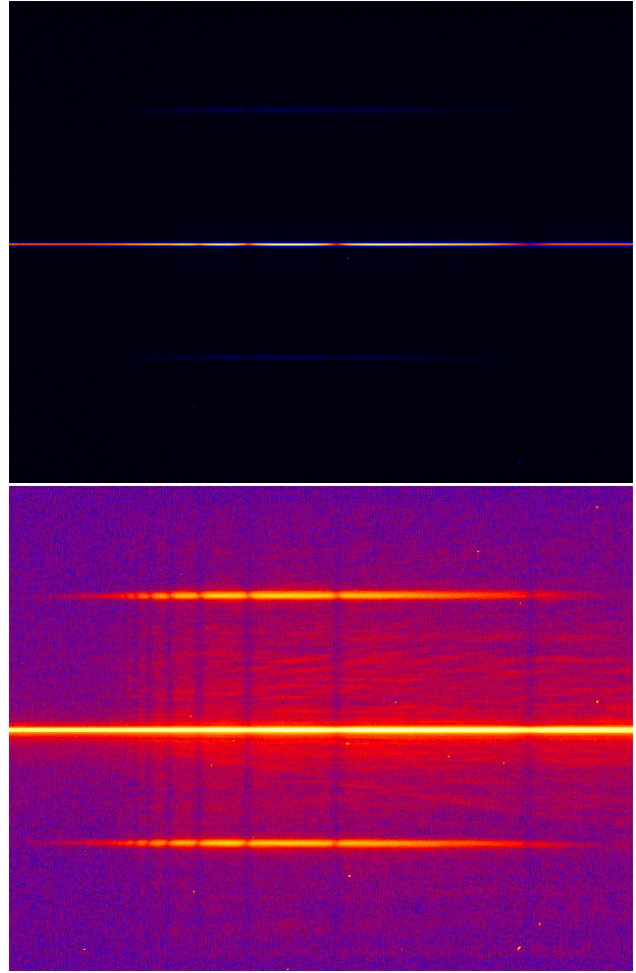

**Figure 2.** (Top) Image of the G430L spectrum of Sirius B (centre), with the spectra of the diffraction spikes of Sirius A (top & bottom) on a linear intensity scale. (Bottom) The same image plotted on a logarithmic scale to enhance the background and scattered light components.

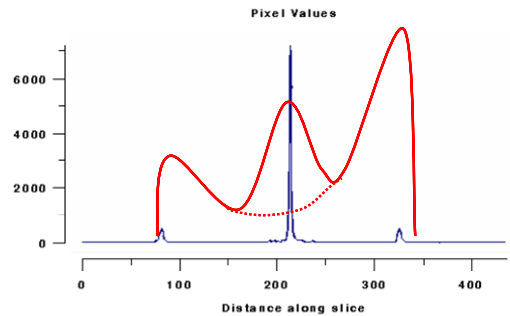

**Figure 3.** Vertical slice through the image in figure 2 showing the relative intensities of the three spectra. For comparison we also show a slice through the spectrum of Kodaira (1967) normalized to the same intensity and spatial scale. The dotted line is an extrapolation of the diffraction spike flux to show the level of contamination in the spectrum of Sirius B.

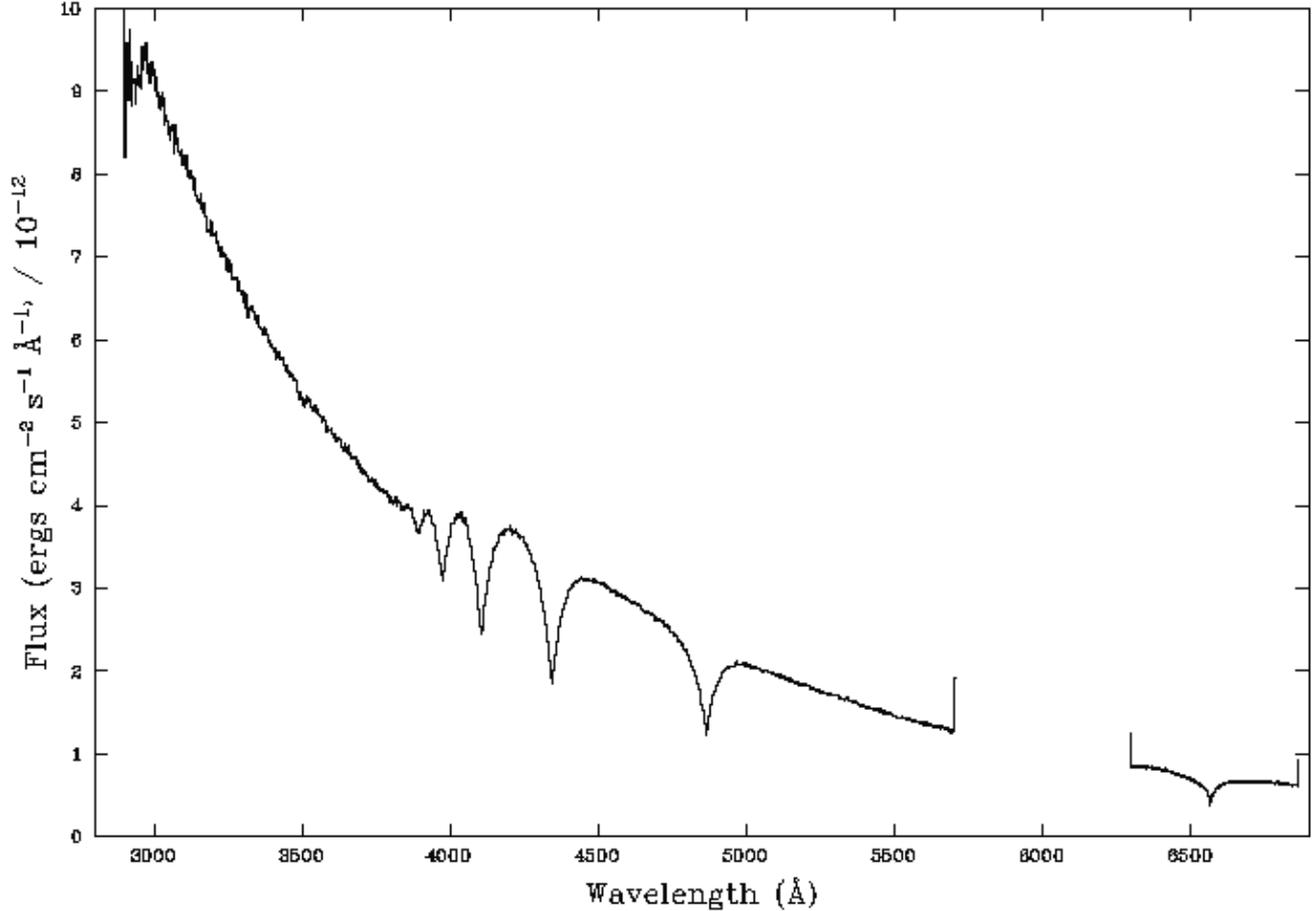

**Figure 4.** Flux calibrated and background subtracted spectra of Sirius B obtained, as described in the text, with the G430L (3000-5700Å) and G750M (6300-6900Å) gratings of the STIS instrument on *HST*.

## 3  ANALYSIS OF THE BALMER LINE SPECTRUM OF SIRIUS B

### 3.1  Spectroscopy

The G430L and G750M spectra of Sirius B were each obtained as a series of 3 exposures to achieve the best possible signal to noise while avoiding saturation of the CCD and to facilitate removal of cosmic rays. These were automatically combined in the standard STIS CCD pipeline before the Sirius B spectrum was extracted, background-subtracted and calibrated (see Kim Quijano et al. 2003). Both resulting spectra are shown in figure 4. It can be seen that the Hα line profile (right) shows a slight "roll-off" in the flux towards short wavelengths. Any light loss from the slit should be wavelength independent, suggesting that this is a calibration artifact. A similar feature is seen in the G750M spectrum of G191-B2B (Proffitt, C.R., 2005, private communication). Therefore, we did not include the Hα profile in the determination of temperature and gravity.

Our standard technique (along with many other authors, see e.g. Bergeron, Saffer & Liebert 1992, Barstow et al. 2003) has been to simultaneously compare the Hβ-Hε lines with synthetic stellar spectra and use a $\chi^2$ fitting technique to determine the best value of $T_{\rm eff}$ and log $g$. These models have been thoroughly described elsewhere (e.g. Barstow et al 2003a, 2003b). To take account of the possible systematic errors inherent in the flux calibration of ground-based spectra, we have usually applied an independent normalization constant to each line. However, in the case of *HST*, there is no atmospheric attenuation to deal with, removing one of the primary uncertainties in the flux calibration process. Furthermore, partly as a result of this, the calibration of the STIS instrument it is extremely accurate and stable. The absolute flux scale is based on four primary DA white dwarf standards (G191-B2B, GD153, GD71 and HZ43), which are pure hydrogen models normalized to Landolt V-band photometry, which yields absolute fluxes determined to ~4% in the far-UV and to ~2% at longer wavelengths (Bohlin 2000; Bohlin Dickinson & Calzetti 2001). Photometric repeatability is in the 0.2-0.4% range. With such accuracy and repeatability, we have fit the complete spectrum of Sirius B covering the Balmer lines from Hβ down to the series limit, a wavelength range from ~5200Å to 3800Å. We used a $\chi^2$ statistic to determine the values of $T_{eff}$ and log $g$ that yield the best agreement between the model and data. The result of this analysis is shown in figure 5, with the best fit values and their associated uncertainties listed in table 2. We note that a standard analysis of the individual lines (discussed above) yields similar results but with larger uncertainties.

The ~0.15% errors on $T_{eff}$ and log $g$ quoted in table 2 are remarkably small and represent the formal internal errors arising from the spectral analysis. However, they only take into account the statistical errors on the data points and do not deal with any systematic errors arising from the analysis procedure, data reduction and or calibration. It is important to question how realistic these are in the light of the magnitudes estimated by other authors. For example, Bergeron et al. (1992) quote typical errors 350K (~1.5-2%) in $T_{eff}$ and 0.05 dex in log $g$ for their sample of DA white dwarfs while Finley et al. (1997) find internal errors of 1% and 0.02 dex respectively. Probably the best and most consistent set of Balmer line analyses is that carried out for the PG sample by Liebert et al. (2005). Their internal uncertainties are 1.2% in $T_{eff}$ and 0.038 dex in log $g$. These are ground-based results. We are operating in uncharted territory as our space-based STIS spectrum is virtually unique in its high signal-to-noise and spectrophotometric fidelity. On the other hand we do not have other observations to help assess the reliability of the quoted errors. Taking a conservative approach we would adopt errors typical of those obtained from the ground-based studies. However, in the context of the systematic uncertainties reported in the latter sections of this paper these spectral analysis errors are small and do not contribute significantly to overall error budget.

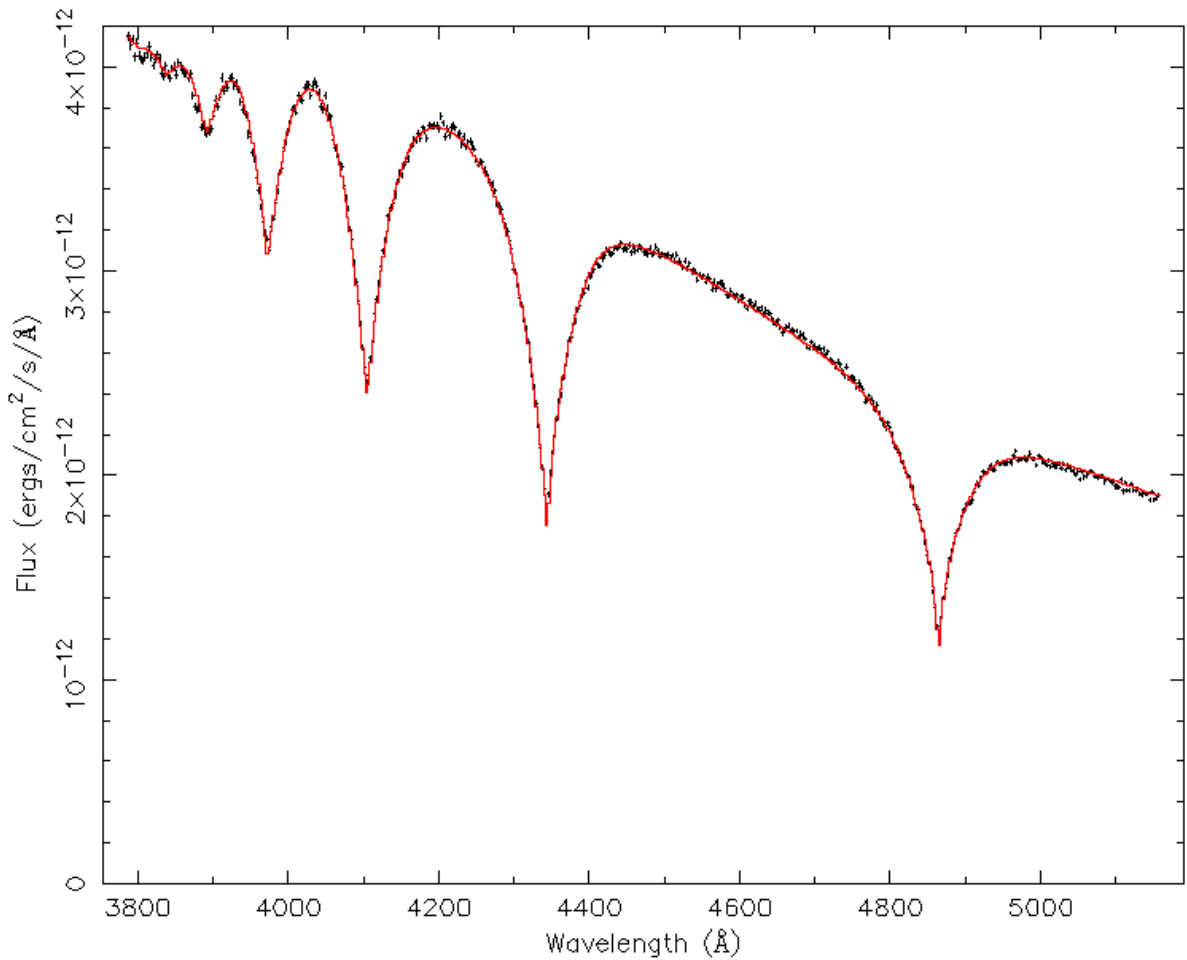

**Figure 5.** Section of the G430L Sirius B spectrum spanning the wavelength range 3800-5200Å (small black crosses, size indicating the statistical errors) with the best-fit synthetic spectrum (red line) corresponding to $T_{eff}$=25,193K and log $g$=8.556.

**Table 2.** Summary of the physical parameters of Sirius B measured or reported by Holberg et al (1998), except for the redshift, which is from Greenstein et al (1971). These values are compared with the most recent results that we have obtained from analysis of the *HST* STIS G430L and G750M spectra. The mass and radius are from the spectroscopic results only and do not take account of the astrometric values.

| Parameter | Value | Error | *HST* Results | |
|---|---|---|---|---|
| $m_V$ | 8.44 | 0.06 | 8.528 | 0.05 |
| $T_{eff}$ (K) | 24,790 | 100 | 25,193 | 37 |
| log $g$ | 8.57 | 0.06 | 8.556 | 0.010 |
| $\pi$ (″) | 0.37921 | 0.00158 | | |
| $V_{gr}$ (km/s) | 89 | 16 | 80.42 | 4.83 |
| $M$ ($M_\odot$) | 0.984 | 0.074 | See table 5 | |
| $R$ ($R_\odot$) | 0.0084 | 0.00025 | See table 5 | |

### 3.2 Photometry

One of the primary sources of uncertainty in the analysis of most previous Sirius B data has been the lack of accurate photometry. Traditionally, this has meant measuring or estimating the V magnitude of the white dwarf. In most past work (eg. Holberg et al 1998), we have used the estimate of V = 8.44 ± 0.06 from Holberg et al (1984), which in turn was based on an average of various published measurements. With the high spectrophotometric accuracy of the *HST* calibration, we have a first opportunity to determine magnitudes for Sirius B in various bands that have an equivalent accuracy to that easily obtained for other white dwarfs where photometric contamination is not and issue.

We can use our observed G430L spectrum to define a model atmosphere spectrum, which accurately represents the absolute Sirius B flux distribution, both in level and in spectroscopic detail. The usefulness of such a model is two fold. First, it covers wavelengths outside the range of the G430L spectrum. Second, it provides a noiseless numerical representation of the absolute spectrum, which can be accurately convolved with various filter functions to give synthetic magnitudes in different bands. The method that we use begins with the new absolute spectrophotometry of Vega (α Lyrae) presented in Bohlin & Gilliland (2004). These authors directly measured the absolute flux of Vega using the STIS CCD and various gratings covering the wavelength range 1700Å to 1μm. A new V magnitude (0.026 ± 0.008) and new absolute flux at 5556Å (3.46 x10$^{-9}$ ergs cm$^{-2}$s$^{-1}$Å$^{-1}$) were also determined. Bohlin & Gilliland estimate the uncertainty in the Vega fluxes to be better than 1%. More importantly, Vega is now directly on the same *HST* White Dwarf flux scale used to define the STIS calibration, and hence our Sirius B fluxes. We can therefore convolve any well-defined set of relative filter functions with the Vega flux distribution to derive appropriate flux zero points for each filter. Convolution of the same filter functions with the Sirius B spectrum, together with the application of the flux zero points, gives the synthetic magnitudes of Sirius B for each filter. Further, it is a simple matter to confirm these procedures by applying the same filter functions and zero points to the set of four fundamental white dwarfs (GD 71, GD 153, HZ 43 and G191B2B), which define the *HST* calibration system and to verify that the resulting synthetic magnitudes match the observed ground-based magnitudes.

In seeking to minimize the effect of scattered light from Sirius A, we chose to use the small 0.2″ slit for both spectroscopic observations. This is also important for obtaining the best possible spectral resolution for the gravitational redshift measurement (see section 3.3). Therefore, although the nominal absolute calibration of the *HST* spectrographs is ~1%, we must also consider the light losses associated with this narrow slit. There are two distinct effects. First, recent adjustments to the flux calibration are not included in the standard pipeline. These are not, as yet, documented formally but have been supplied to us by C.R. Proffitt (2005, private communication). For the G430L grating and 52x0.2″ aperture, the flux measurement from the pipeline is overestimated by a smooth wavelength dependent factor ranging from ~5-7%. Accordingly, we have corrected the measured fluxes used in the analyses reported here using the appropriate function. There is insufficient calibration data available for a similar analysis to be carried out for the G750M 52x0.2″ aperture combination.

While light losses, arising from the use of the small aperture are in principle taken into account in the calibration pipeline, there is an enhanced scatter in the accuracy of the flux determination due to stochastic effects on the placement of the source in the slit. Bohlin (1998) estimates this to have an rms value of 4.5% for the 52x0.2″ slit, dominating the 1% absolute calibration error, which yields a 0.05 magnitude uncertainty in the phototmetry. Therefore, the uncertainty that we can assign to the synthetic magnitudes and fluxes measured here will be correspondingly greater than the formal absolute error. However, we note that the increased uncertainty associated with light loss in the slit can only be single sided, as flux can only be lost and not gained relative to the true brightness. Therefore, any determination of the flux, and the corresponding magnitudes, has an asymmetric range of uncertainty of +5% and -1%.

We have chosen the UBVRI filters responses defined by Cohen et al. (2003) for this work. These include detailed atmospheric transition modifications to the Landolt (1992) UBVRI filter functions. Bohlin & Gilliland use the Cohen et al. V band filter function to establish their V magnitude for Vega. In addition to the Cohen et al. UBVRI filter functions; we also use the observed V magnitudes of Vega from Bessel, Castelli & Plez (1998), except for V where we use the Bohlin & Gilliland Vega V magnitude. The relation between integrated fluxes, $F_{int}$, the Vega Magnitudes, Filter Constants is given in equations P1 and P2, where $f(\lambda)$ is the Vega flux and $S(\lambda)$ the relative filter response function. In Table 3 we provide the results of these calculations for Sirius B, the UBVRI filters; including the logs of the flux zero points, the observed Vega magnitudes, and synthetic Sirius magnitudes for each filter.

$$F_{int} = \frac{\int f(\lambda)S(\lambda)d\lambda}{\int S(\lambda)d\lambda}$$

Filter Const. = Vega Mag. + 2.5log($F_{int}$)

Using these same procedures for GD 71, GD 153, HZ 43 and G191 B2B, the absolute values of the differences between the synthetic and the observed (Landolt) magnitudes are less than 0.007 magnitudes for each filter, except for the U band. The numerical definition of the U filter function extends shortward of the atmospheric cutoff and thus will not correspond to any observed U magnitude. Our synthetic Sirius B U magnitude therefore represents a hypothetical observation above the earth's atmosphere rather than any realizable ground-based observation. Because our calculations are self-consistent with respect to the *HST* flux scale, we have set the uncertainties for each magnitude to 0.05 magnitudes to accommodate the estimated uncertainty in that scale and to take account of the small, 0.2″, aperture used in the observations. This is not much smaller than the uncertainty obtained by Rakos and Havlen (1977). Using V = 8.528 ± 0.05 and our adopted trigonometric parallax for Sirius B, we find an absolute magnitude of $M_v$ = 11.427 ± 0.05.

**Table 3.** Synthetic Photometry of Sirius B

| Band | Vega Mag. | Filter Const. | Sirius B Mag. |
|------|-----------|---------------|---------------|
| U | 0.0240 | -21.0020 | 7.256 + 0.01/-0.05 |
| B | 0.0280 | -20.4477 | 8.394 + 0.01/-0.05 |
| V | 0.0260 | -21.0503 | 8.528 + 0.01/-0.05 |
| R | 0.0370 | -21.6061 | 8.656 + 0.01/-0.05 |
| I | 0.0330 | -22.3704 | 8.802 + 0.01/-0.05 |

### 3.3 The gravitational redshift

The primary aim of obtaining the higher resolution G750M spectrum was to use the narrow Hα core to obtain a gravitational redshift for Sirius B. To do this we cross-correlated a synthetic Hα profile, computed for the temperature and gravity determined for the other Balmer lines, with the observed line and calculated the relative Doppler shift between the two using a $\chi^2$ minimisation technique. With the narrow Hα core, this technique yields a formal fractional uncertainty in $z$ of ~1%.

With such precision, it is important to consider any systematic effects that might contribute to the overall uncertainty. For example, with a predicted redshift of ~70km/s (before correction for γ and K velocities), the observed wavelength shift will be ~1.5Å. Therefore it is necessary to use a reference wavelength for Hα that is accurate to better than 0.01Å. Accordingly, we have recalculated the wavelength of Hα from the weighted mean of the fine structure energy levels, obtaining a value of 6564.6271Å in vacuum, which we adopt for this analysis. We note that, although we are dealing with visible light wavelengths for which the wavelength calibration is usually carried out in air (for ground-based telescopes) the STIS calibration refers to vacuum at all wavelengths.

Although we take account of Stark broadening in calculating the profiles of the Balmer (and Lyman) lines in the synthetic spectra, using the tables of Lemke (1997), we have not routinely considered the possible Stark shifts of the lines. These are predicted to be quite small but are always in the redward direction and could contribute to an increased redshift measurement compared to the true gravitational value. Greenstein et al. (1971) estimated such a shift to be about 8 km/s, based on the data of Wiese and Kelleher (1971). In the context of the ± 16 km/s uncertainty in their redshift measurement, this is not significant and the magnitude of the Stark shift was not considered further, due to the noise in the Wiese and Kelleher (1971) data for small shifts and the complexity of the radiative transfer problem. However, compared to our redshift measurements, which have a formal 1% accuracy, this ~10% contribution to the measured redshift (equivalent to ~0.15Å in the measured wavelength shift) would be extremely important. However, the reported possible shifts are highly dependent on the plasma density which is determined by the Balmer line formation depth. A full radiative transfer treatment is needed to determine whether or not the Stark pressure shift makes a significant contribution to the measured redshift. Consequently, we have modified the spectral synthesis programe SYNSPEC to include the effect of Stark shifts in the calculation of the Balmer and Lyman line profiles. We followed the procedure outlined by Grabowski et al. (1987). For a white dwarf of the temperature and gravity of Sirius B, the additional Stark shift predicted by the new spectral synthesis calculations is tiny and does not need to be considered in this (or any) analysis.

The H$\alpha$ line extends across ~300 pixels of the G750M spectrum. It is possible to use just the narrow line core or the core and wings for the redshift measurement. It is debatable whether using the narrow core alone or making use of the additional information available in the line wings is the best approach. Table 4 records the observed wavelength shift and corresponding velocity for several different pixel ranges. There is a significant scatter in the values obtained, compared to the formal statistical errors, indicating the possible level of any systematic errors. In our further analysis, we adopt the mean of these redshift values and their standard deviation as an indication of the true uncertainty in the measurement. In addition, we must also consider the uncertainty in the calibration of the wavelength scale, which is ~0.2 pixels (0.12Å) for the narrow slit, giving a total uncertainty of 6% of the measured redshift.

**Table 4.** Redshift measurements made for different pixel ranges centred on the wavelength of the H$\alpha$ line.

| Pixel no | z | $\Delta\lambda$ | v(km s$^{-1}$) |
|---|---|---|---|
| 14 | 2.4083x10$^{-4}$ | 1.581 | 71.82 |
| 114 | 2.4404x10$^{-4}$ | 1.602 | 72.45 |
| 214 | 2.2500x10$^{-4}$ | 1.477 | 68.71 |
| 314 | 2.2332x10$^{-4}$ | 1.466 | 68.35 |
| Mean | | | 70.33±1.82 |

To convert this into a gravitational redshift it is necessary to take account of the radial velocity of Sirius B, which consists of the gamma velocity of the system and the K velocity of Sirius B. The apparent velocity is the algebraic sum of the gravitational redshift, the orbital velocity of Sirius B and the $\gamma$ velocity of the system barycenter. We have independently determined the latter two quantities in order to estimate the intrinsic gravitational redshift of the white dwarf (Holberg 2005). The system velocity or $\gamma$ velocity can be directly determined from the apparent radial velocity of Sirius A as a function of time. Holberg (2005) has fit the published velocities of Sirius A allowing the constant $\gamma$ velocity to vary as a free parameter. The observed velocities were taken from the literature and included early, turn of the last century, photographic measurements as well as more recent CCD observations from nine observatories. In most instances the data points represented annual or semi-annual averages from a given observatory. Where uncertainties were not provided they were estimated from the standard deviations of the means or were assigned a value as typical for a given observatory at that period. No attempt was made to adjust velocity zero points between different observatories. Several obviously discrepant observations were excluded. The direct $\chi^2$ fit for the $\gamma$ velocity yields -7.85 ± 0.72 km s$^{-1}$. The result is only very weakly dependent on any reasonable selection of orbital parameters and stellar mass ratio and is in good agreement with earlier results obtained by Campbell (1905), Aitken (1918) and van den Bos (1960); -7.4 km s$^{-1}$, -7.37 km s$^{-1}$, and -7.43 km s$^{-1}$, respectively. None of these earlier results, however, appear to have corrected for the gravitational redshift of Sirius A (+0.75 km s$^{-1}$). When this correction is applied, the resulting value of $\gamma$ = -8.60 ± 0.72 km s$^{-1}$. The orbital Doppler velocity of Sirius B on 2004 Feb 6 (-1.49 km s$^{-1}$) can be obtained directly from the work of Holberg (2005). Thus, the additive correction to be applied to our apparent Doppler velocity of Sirius B is -10.09 ± 0.72 km s$^{-1}$, yielding a value of 80.42 ± 4.83 for the gravitational redshift of the white dwarf.

**4 DISCUSSION**

We have presented an initial analysis of the first Balmer line spectrum of Sirius B obtained from space. While this is not the only Balmer line spectrum obtained, it is certainly the only one to have eliminated the problem of the scattered light from Sirius A, providing a clean background subtracted spectrum from which accurate determinations of $T_{eff}$, log $g$ and the gravitational redshift can be made. It is clear, from tables 2 and 5, that the uncertainties in the determination of all these parameters are considerably improved from their earlier values. Within the quoted errors, the values obtained in this work are mostly compatible with the previous determinations, apart from $T_{eff}$. However, it is important to note that the older measurements of $T_{eff}$ and log $g$ were made using a previous generation of stellar atmosphere calculations, which might explain the difference between the temperature values. A thorough analysis of other data sets with the most recent models will be required to resolve this.

With improved measurements of the physical parameters of Sirius B it should now be possible to improve the accuracy of the determination of the mass and radius and, as a result, provide a more definitive test of the white dwarf mass radius relation.

**4.1 The radius and mass of Sirius B**

Rather than calculating the photometric radius of Sirius B from the synthetic V magnitude, it is more straightforward to simply use the normalization applied to the best fit spectral model to match the observational data. This conveniently defines the stellar solid angle (R$^2$/D$^2$).

Using our adopted trigonometric parallax we can then calculate the white dwarf radius directly. We have performed this exercise for both our G430L and G750M data, as summarized in table 5. The observational uncertainties in R are dominated by the systematic errors in the *HST* flux scale. It can be seen that the values of $R^2/D^2$ and R are not identical for each grating, although they do agree within the identified systematic uncertainties. We note that the G430L flux scale has had a flux scale correction applied that is not included in the pipeline processing (as discussed in section 3.2). No similar correction could be applied to the G750M data because the relevant calibration information is not available.

Having determined the radius of Sirius B, there are two independent ways that we can estimate its mass from the available data, using relations including the surface gravity or the gravitational redshift:

$$g = GM/R^2 \qquad V_{gr} = 0.636 M/R$$

Since there are two estimates for the stellar radius, we have made separate calculations of the mass using each of these (table 5).

**Table 5.** The mass and radius of Sirius B calculated for the different values of R related to the normalization constant determined for each of the gratings used.

| Grating | G430L | G750M |
|---|---|---|
| $R^2/D^2$ | $4.662 \times 10^{-21}$ | $4.996 \times 10^{-21}$ |
| $R_\odot$ (×10$^{-3}$) | 8.004 + 0.372 / -0.081 | 8.330 + 0.383 / -0.083 |
| $M_\odot$ (g) | 0.841 + 0.080 / -0.026 | 0.911 + 0.084 / -0.027 |
| $M_\odot$ ($V_{gr}$) | 1.012 ± 0.060 | 1.050 ± 0.063 |

It is clear that the two different methods of estimating the white dwarf mass are giving us different answers. However, within the overall uncertainties that we have uncovered in the flux and wavelength calibration, the results are formally more or less compatible but at the very extremes of the range within which they remain consistent. Indeed the 1σ ranges of *M* determined for the radius derived from the G430L flux level do not quite overlap. This is illustrated in figure 6. A method for combining all the previously available data has been described by Holberg et al (1998) and we compare that analysis with the new values of *M* and *R* listed in table 5 in Figure 6. This shows 1σ and 2σ error regions determined by Holberg et al (1998) together with the allowed astrometric mass range from Gatewood and Gatewood (1978). Since we have no information on which of the G430L or G750M flux levels yields the "best" determination of the white dwarf radius we plot both sets of results. The upper error bars are those corresponding to the G750M measurement with the mass derived from the surface gravity on the left and that from the gravitational redshift on the right (blue and red crosses respectively. The lower pair of error bars (green and purple) are the corresponding mass determinations for the lower radius obtained from the G430L results.

Although the different determinations of M are not significantly different outside the assigned 1σ uncertainties ranges, it is a concern that we do not get better agreement between the various methods of determining *M*. In theory it should be possible to treat the data we have obtained with *HST* in the manner described by Holberg et al (1998) to determine a "best fit" mass. However, if we do this with our results, the $\chi^2$ minimization technique is rather unstable and very sensitive to the adopted 1σ errors, the values of which are themselves uncertain in the light of the systematic effects we have discussed in this paper. Therefore, we do not believe that the Holberg et al. (1998) method will give a reliable result with the current data.

Comparing the values of *M* and *R* obtained in this work with the earlier results is interesting. Individually, only the G430L/surface gravity measurement of *M* is incompatible with Holberg et al. (1998). However, if the astrometric mass determination is taken into account, only the values of *M* derived from the gravitational redshift are consistent. Of the measurements we have made, the most problematic is the determination of the stellar radius. Using the narrow slit, it is possible that there could be a loss of light and, as a result, an erroneously low flux determination. This would lead to a low estimate for the white dwarf radius. We have tried to take account of this possibility in the assignment of experimental errors. However, if there really were a light loss at the ~5% level (the extreme possibility) the overall agreement between the mass determinations would be improved. An independent measurement of the white dwarf flux would be helpful in resolving this question.

Although the main theme of this analysis has been to obtain measurements of *M* and *R* that are independent of the white dwarf evolutionary models, given the difficulty in obtaining a reliable stellar radius (on which our measurements of mass dependent), it is worthwhile carrying out an empirical determination of *M* using the them. Taking the values of $T_{eff}$ and log *g* determined in this paper, we interpolate between the evolutionary models of Wood (1995), which are computed for discrete white dwarf masses in steps of 0.2$M_\odot$, obtaining *M* = 0.978 ± 0.005 $M_\odot$ and *R* = 0.00864 ± 0.00012 $R_\odot$ for Sirius B. The corresponding gravitational red-shift is 72.0 ± 1.0 km s$^{-1}$. We note that, as in the discussion in section 3.1 on the spectral analysis, the formal errors quoted here are probably factors ~5-10 too small if we adopt errors for $T_{eff}$ and log *g* typical of the ground-based studies. With this in mind, these empirical values of *M* and *R* are consistent with those obtained directly from the STIS observations, provided we adopt the flux level taken from the G750M observation rather than the more thoroughly calibrated G430M.

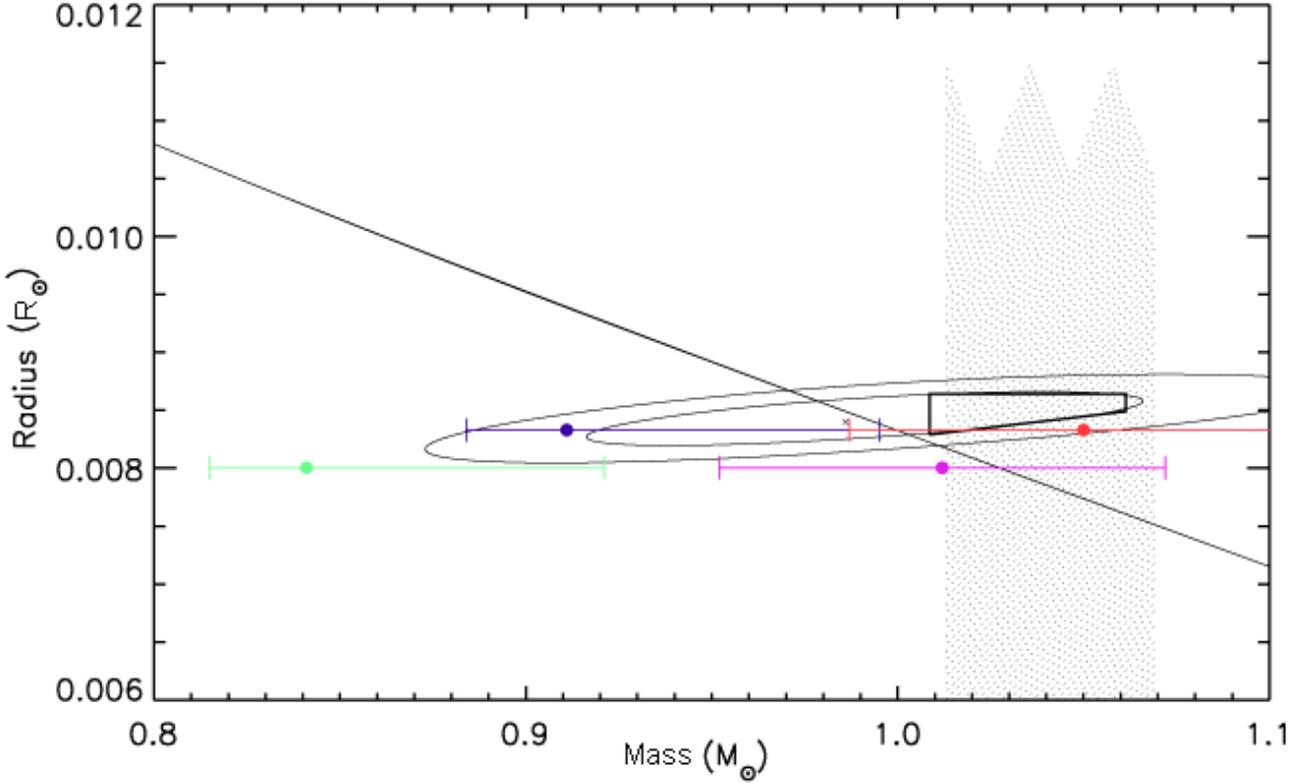

**Figure 6.** Sirius B mass and radius in solar units compared to the mass-radius relation for carbon-core white dwarfs. The two ellipsoidal contours are the 1σ and 2σ regions determined by Holberg et al (1998). The vertical stippled region is the range of the astrometric mass determined by Gatewood and Gatewood (1978) while the heavy trapezoid is the joint 1σ spectroscopic and astrometric eastimate of Holberg et al (1998). The sloping black curve is the theoretical mass-radius relation for a ~25000K DA white dwarf with a C-O core (Wood 1995). The upper error bars are those corresponding to the G750M measurement with the mass derived from the surface gravity on the left and that from the gravitational red-shift on the right (blue and red crosses respectively). The lower pair of error bars (green and purple) are the corresponding mass determinations for the lower radius obtained from the G430L results.

The evolutionary calculations of mass and radius can also be compared to our new measurement of the gravitational red-shift, since this gives a value (126.4 ± 7.6) for *M/R* directly. Matching this to the relation of Wood (1995) shown in figure 6 yields $M = 1.02 \pm 0.02$ and $R = 0.0081 \pm 0.0002$. Importantly, these estimates are not dependent on the model atmosphere calculations, unlike the values derived from the $T_{eff}$ and log *g* measurements, and probably have a more robust error determination. Within the overall uncertainties, there is internal consistency between the various methods we have discussed for obtaining *M* and *R*. They are also in agreement with the earlier study of Holberg et al. (1998).

## 5  CONCLUSION

We have obtained an exquisite spectrum of the complete Balmer line series for Sirius B. This is the first such spectrum to be acquired, apart from old ground-based photographic spectra, and can be used to provide an important determination of the stellar temperature ($T_{eff}$ = 25193 ± 37 K) and gravity (log *g* = 8.566 ± 0.010). In addition we have obtained a new, more accurate, gravitational red-shift of 80.42 ± 4.83 km s$^{-1}$ for Sirius B. Combining these results with the photometric information available in our spectra and the *Hipparcos* parallax we have provided new determinations of the stellar mass and radius for comparison with the theoretical mass-radius relation. However, there are some disparities between the values of stellar mass obtained by two different routes and we have identified significant systematic uncertainties that make the observational errors larger than we had hoped. While we have attempted to makes measurements of the mass and radius of white dwarf independently of the evolutionary model calculations (e.g. Wood 1992, 1995), we get much better agreement between our results and those of other authors if we use our spectroscopic measurements of $T_{eff}$ and log *g* in conjunction with the theoretical mass-radius relation. Our best estimates of *M* and *R* from this approach are 0.978 ± 0.005 M$_\odot$  and $0.00864 \pm 0.00012$ R$_\odot$ respectively. The gravitational redshift gives us an estimate of *M/R* directly, which can also be compared to the Wood (1995) models, yielding $M = 1.02 \pm 0.02$ and $R = 0.0081 \pm 0.0002$. These values are all in good agreement with the measurements of other authors and internally consistent with our independent measurements, provided we utilize the stellar absolute flux obtained from the G750M grating.

A particular problem we have encountered is that of possible light loss due to the use of the narrow 50x0.2″ slit, yielding a measured flux for Sirius B lower than the true value. Indeed, the better consistency of results derived from the G750M flux compared to the better calibrated G430L data is indicative of a problem. We anticipate that it will be possible to improve on the measurement derived from our STIS spectra in the future, by making use of the WFPC2 images we have acquired for the study of the binary orbit. In addition, with a wealth of other data also available from soft X-ray, EUV and far-UV wavebands, an important exercise will be to combine all the information we have to provide the best possible estimate of the mass and radius of Sirius B.

**Acknowledgements.** These observations were obtained through the Guest Observer programme of the Hubble Space Telescope. We are grateful to the support astronomers involved for their assistance in scheduling this difficult observation. We would particularly like to thank Charles Proffitt for his assistance in understanding the calibration issues for these spectra and for providing important information that is not yet available in the public domain. MAB and MRB acknowledge the support of the Particle Physics and Astronomy Research Council, UK. JBH wishes to acknowledge support from STScI grant GO 09762.